\documentclass[12pt]{article}
\usepackage{graphicx,cite}

\input{epsf}

\newcommand{\nc}{\newcommand}
\nc{\beq}{\begin{equation}}
\nc{\eeq}{\end{equation}}
\nc{\beqa}{\begin{eqnarray}}
\nc{\eeqa}{\end{eqnarray}}

\def\gsim{\mathrel{\rlap{\lower4pt\hbox{\hskip1pt$\sim$}}
    \raise1pt\hbox{$>$}}}       

\begin{document}



\title{\large{\bf Cosmological Constant and Noncommutative Spacetime}}

\author{Xavier~Calmet \\
Universit\'e Libre de Bruxelles \\ 
Service de Physique Th\'eorique, CP225 \\ 
Boulevard du Triomphe  (Campus plaine)\\
B-1050 Brussels, Belgium
\\
email: xcalmet@ulb.ac.be}

\date{October, 2005}

\maketitle

\begin{abstract}
We show  that the cosmological constant appears as a Lagrange multiplier if nature is described by a canonical noncommutative spacetime. It is thus an arbitrary parameter unrelated to the action and thus to vacuum fluctuations. The noncommutative algebra restricts general coordinate transformations to four-volume preserving noncommutative coordinate transformations. The noncommutative gravitational action is thus an unimodular noncommutative gravity. We show that spacetime noncommutativity provides a very natural justification to an unimodular gravity solution to the cosmological problem. We obtain the right order of magnitude for the critical energy density of the universe if we assume that the scale for spacetime noncommutativity is the Planck scale.
\end{abstract}


\newpage
There are strong observational evidences that the universe is dominated by dark energy which accounts for roughly 70$\%$ of its energy. A natural candidate for dark energy is a cosmological constant. However, from a high energy theorist point of view, see e.g. \cite{Weinberg:1988cp} for a review, it is difficult to understand why this constant is so small and not of the order of the fourth power of the TeV scale  or maybe even of the forth power of the Planck scale. The scale depends on assumptions on what kind of physics lies ahead of the standard model of particle physics. A naive, maybe too naive, estimate of the contribution to the cosmological constant of quantum fluctuations of the fields of the standard model yields a result which is off by easily 100 orders of magnitude from the value observed in nature. The cosmological problem can actually be viewed as two problems. The first one is why is the cosmological constant not of the order of the Planck scale. The second one is why is it small and not vanishing. There are many attempts to resolve these issues \cite{Weinberg:1988cp}, however  none of them is really satisfactory.

In this work we shall reconsider an old approach to the cosmological constant problem based on unimodular gravity\cite{Weinberg:1988cp,UNI,vanbij,Henneaux:1984ji,wilczek,buchmuller,henneaux,unruh}. Because unimodular gravity arises from an ad hoc restriction of general coordinate transformations to those preserving the four-volume, i.e. the determinant of the metric to be constant, it does not solve the cosmological constant problem, but nevertheless it allows to rephrase the problem since in that framework the cosmological constant appears as a Lagrange multiplier and is arbitrary, i.e. independent of the parameters of the action and in particular  of the vacuum expectation values of the scalar fields of the particle physics sector of the model. However, we shall argue that unimodular gravity is not a choice if spacetime is noncommutative. In that case we find solutions to the cosmological constant problems which imply a relation between very short distance physics, potentially the Planck length, and the cosmological constant i.e. the size of the universe.
The solution to the first problem comes from the fact that on a noncommutative spacetime, the cosmological constant is arbitrary as it appears as a Lagrange multiplier and is thus independent of the parameters of the action. This is a rigorous result and a consequence of a noncommutative spacetime. The solution to the second problem is similar to the one proposed in \cite{Ng:1999tx,Ng:2003jk,Ahmed:2002mj,LamPred,sorkin2} but motivated by arguments coming from noncommtutative geometry. However, it is based on quantum gravity arguments and is therefore not as rigorous as the solution to the first problem.

There are many indications coming from thought experiments which are trying to unify General Relativity and Quantum Mechanics that spacetime may be discrete rather than
continuous \cite{minlength1,Padmanabhan:au,Garay:1994en}.  It has recently been shown \cite{Calmet:2004mp,Calmet:2005mh} that
no macroscopic experiment can be sensitive to discreteness of position
on scales less than the Planck length. Any device subject to classical General Relativity, Quantum Mechanics and causality capable of such resolution would be so massive that it
would have already collapsed into a black hole.

A way to implement the notion of a minimal length in quantum field theory and in classical General Relativity is to consider these theories  on a canonical noncommutative spacetime. Positing a noncommutative relation between e.g. $x$ and $y$ implies  $\Delta x \Delta y \geq l^2$. This also implies that a spacetime volume is quantized $\Delta V \geq l^4$,  e.g. the area in  the 2d Euclidean space cannot be smaller than $l^2$ or $(\Delta x)^2+(\Delta y)^2\geq l^2$ which is the Euclidean distance,  or ``radius" of the area.

In this letter we propose a solution to the cosmological problem based on an unimodular theory of gravity. As recently shown \cite{Calmet:2005qm}, the only general coordinate transformations $\xi^\mu(\hat x)$ that leave the canonical noncommutative algebra invariant
\begin{eqnarray} \label{a}
[ \hat x^\mu, \hat x^\nu ]=i \theta^{\mu \nu},
\end{eqnarray}
where $\theta^{\mu\nu}$ is constant and antisymmetric, are of the form:
$\xi^{\mu}(\hat x)=\theta^{\mu\nu} \partial_{\nu} f(\hat x)$.  We now give a proof of  this result. The commutator (\ref{a}) explicitly violates general coordinate covariance since $\theta^{\mu\nu}$ is constant in all reference frames. However, we can identify a subclass of general coordinate transformations,
\begin{equation}
\hat x^{\mu \prime}=\hat x^{\mu}+\xi^{\mu}(\hat x),
\label{c}
\end{equation}
which leave the algebra (\ref{a}) invariant. Under the change of coordinates (\ref{c}) the commutator (\ref{a}) transforms as:
\begin{eqnarray}
[\hat x^{\mu \prime}, \hat x^{\nu \prime}]&=&
\hat x^{\mu \prime}\hat x^{\nu \prime}- \hat x^{\nu \prime} \hat x^{\mu \prime} 
=i \theta^{\mu\nu} + [\hat x^\mu,  \xi^\nu] + [ \xi^\mu, \hat x^\nu] + {\cal O}(\xi^2).
\label{d}
\end{eqnarray} 
Requiring that $\theta^{\mu\nu}$ remains constant yields the following partial differential equations:
\begin{eqnarray}
\theta^{\mu \alpha}  \partial_{\alpha}  \xi^\nu(\hat x) = \theta^{\nu \beta}  \partial_{\beta} \xi^\mu(\hat x).
\end{eqnarray}
The most general solution to these partial differential equations can be easily found:
\begin{equation} \label{nctransf}
\xi^{\mu}(\hat x)=\theta^{\mu\nu} \partial_{\nu} f(\hat x),
\label{e}
\end{equation}
where $f(\hat x)$ is an arbitrary field.  The Jacobian of these restricted coordinate transformations is equal to one. This implies that the four-volume element is invariant: $d^{4}x^{\prime}=d^{4}x$. These noncommutative transformations correspond to volume preserving diffeomorphisms which preserve the noncommutative algebra. Volume-preserving transformations have been previously discussed in \cite{Jackiw:2002pn,Lukierski:2002ew} in different physical frameworks. A canonical noncommutative spacetime thus restricts general coordinate transformations to volume preserving coordinate transformations. These transformations are the only coordinate transformations that leave the canonical noncommutative algebra invariant.  They form a subgroup of the unimodular transformations of a classical spacetime. The relation (\ref{nctransf}) nevertheless does not introduce further relations between the $\xi^{\mu}(\hat x)$ and there are thus three independent transformations and no further constraint on the metric besides having a constant determinant.
It is interesting to study the limit $\theta^{\mu\nu} \to 0$, in that case the constraint (\ref{a}) disappears and general coordinate transformations can be considered. This seems like a triviality but it will have some importance when it comes to formulate an action for gravity on a noncommutative spacetime.

The version of General Relativity based on volume-preserving diffeomorphism is known as the unimodular theory of gravitation \cite{UNI} (see \cite{vanbij,Henneaux:1984ji,wilczek,buchmuller,henneaux,unruh} for more recent works on unimodular gravity). 
Unimodular gravity here appears as  a direct consequence of spacetime noncommutativity defined by a constant antisymmetric $\theta^{\mu\nu}$. One way to formulate gravity on a noncommutative spacetime has been presented in \cite{Calmet:2005qm}. Our approach might not be unique, but if the noncommutative model is reasonable, it must have a limit in which one recovers usual general relativity in the limit in which $\theta^{\mu\nu}$ goes to zero. For small $\theta^{\mu\nu}$ we thus expect
\begin{eqnarray} \label{NCaction}
S_{NC}= \frac{-1}{16 \pi G} \int d^4x \sqrt{-g} R(g^{\mu\nu}) + {\cal O}(\theta),
\end{eqnarray}
where $R(g^{\mu\nu})$ is the usual Ricci scalar and $g$ is the determinant of the metric. If we restrict ourselves to the transformations (\ref{nctransf}), the determinant of the metric is always equal to minus one, the term $\sqrt{-g}$ in the action is thus trivial. However, as mentioned previously, we recover full general coordinate invariance in the limit $\theta^{\mu\nu}$ to zero and it is thus important to write this term explicitly to study the symmetries of the action. In order to obtain the equations of motion corresponding to this action, we need to consider variations of (\ref{NCaction}) that preserve 
$g=\mbox{det} g^{\mu\nu}=-1$, i.e. not all the components of $g_{\mu\nu}$ are independent. One thus introduces a new variable \cite{unruh,henneaux} $\tilde{g}_{\mu\nu}=g^\frac{1}{4}  g_{\mu\nu}$, which has explicitly a determinant equal to one.

The equations of motion are:
\begin{eqnarray} \label{eqofmo}
R^{\mu\nu}-\frac{1}{4} g^{\mu\nu}R=-8 \pi G (T^{\mu\nu}-\frac{1}{4} g^{\mu\nu} T^\lambda_{\ \lambda})+ {\cal O}(\theta).
\end{eqnarray}
These equations do not involve a cosmological constant and the contribution  of vacuum  fluctuations automatically cancel on the right-hand side of eq.(\ref{eqofmo}). Spacetime noncommutativity because it imposes an unimodular theory of gravity thus provides a mechanism to cancel the vacuum  fluctuation contributions to the cosmological constant. It can be shown that the differential equations (\ref{eqofmo})  imply
\begin{eqnarray} \label{einstein}
R^{\mu\nu}-\frac{1}{2} g^{\mu\nu}R-\Lambda g^{\mu\nu}=-8 \pi G T^{\mu\nu}-{\cal O}(\theta),
\end{eqnarray}
i.e. Einstein's equations  \cite{Einstein:1916vd} of General Relativity with a cosmological constant $\Lambda$ that appears as an integration constant and is thus uncorrelated to any of the parameters of the action (\ref{NCaction}).  To derive eq.(\ref{einstein}) from eq.(\ref{eqofmo})
one needs to impose energy conservation and the Bianchi identities. Because any solution of Einstein's equations with a cosmological constant can, at least over any topologically $R^4$ open subset of spacetime, be written in a coordinate system with  $g=-1$, the physical content of unimodular gravity is identical at the classical level to that of Einstein's gravity with some cosmological constant \cite{unruh}.

As stressed by Weinberg \cite{Weinberg:1988cp}, unimodular gravity on a commutative spacetime does not represent a solution to the cosmological constant problem, since it is a matter of choice to fix $\mbox{det} g_{\mu\nu}=-1$. On a noncommutative spacetime, this is not a choice anymore, the symmetries of the noncommutative algebra restrict the symmetries of the gravitational action to unimodular  coordinate transformations. This noncommutative algebra excludes any connection between the cosmological constant and the parameters of the action and in particular the Planck scale. The cosmological constant is an arbitrary parameter and can thus naturally have any value and it is not fixed by any of the scales appearing in the matter action. 

The connection between the cosmological constant and spacetime noncommutativity relies on the symmetries of the classical noncommutative gravitational action. In the sequel we shall give an estimate of the value of the cosmological constant on a noncommutative spacetime. We shall make use of arguments based on quantum unimodular gravity. These arguments are thus less rigorous than in the first part of the paper.

We now come to the second cosmological problem and rephrase the arguments developed in \cite{Ng:1999tx,Ng:2003jk,Ahmed:2002mj,LamPred,sorkin2} within the framework of noncommutative gravity. It has been shown that the quantization of an unimodular gravity action proposed by Henneaux and Teitelboim \cite{henneaux}, which is an extension of the action defined in eq. (\ref{NCaction}),  leads to an uncertainty relation between the fluctuations of the  volume $V$ and those of the cosmological constant $\Lambda$: $\delta V \delta \Lambda\sim 1$ using natural units, i.e. $\hbar=l_p=c=m_p=1$. Now if spacetime is quantized, as it is the case for noncommuting coordinates, we expect the number of cells of spacetime to fluctuate according to a Poisson distribution, $\delta N \sim \sqrt{N}$, where $N$ is the number of cells. This is however obviously an assumption which could only be justified by a complete understanding of noncommutative quantum gravity. It is then natural to assume that the volume fluctuates with the number of spacetime cells $\delta V=\delta N$. One finds $\delta V \sim \sqrt{V}$ and thus $\Lambda \sim \frac{1}{\sqrt V}$, i.e., we recover the result discussed in \cite{Ng:1999tx,Ng:2003jk,Ahmed:2002mj}: $\Lambda \sim V^{-\frac{1}{2}}$ or in other words the cosmological constant of a noncommutative spacetime contributes to an energy density $\rho$
\begin{eqnarray} \label{crit}
\rho \sim \frac{1}{R_H^2}\sim H_0^2=\rho_{\mbox{critical}}, 
\end{eqnarray}
which is of the order of the critical energy density observed in nature. Since the effects of spacetime noncommutativity are assumed to be weak, we have the usual relation between the Hubble radius and the volume of spacetime. Here we assume that the scale for the quantization of spacetime is the Planck scale. Clearly the derivation of eq.(\ref{crit}) assumes that quantum fluctuations are important and a rigorous derivation would probably require a nonperturbative calculation. It is however remarkable to obtain the right order of magnitude based on simple arguments motivated by a quantized spacetime. A crucial assumption made in \cite{Ng:1999tx,Ng:2003jk,Ahmed:2002mj} as well is that the value of cosmological constant fluctuates around zero. This was made plausible by Baum \cite{Baum:1984mc} and Hawking \cite{Hawking:1984hk} using an Euclidean formulation of quantum gravity. From our point of view this is the weakest point of the proposal developed in  \cite{Ng:1999tx,Ng:2003jk,Ahmed:2002mj} and classical, i.e. non-quantized, noncommutative gravity has nothing new to add to this question. 

The phenomenology of the relation (\ref{crit})  has been discussed in  \cite{Ahmed:2002mj} where it is made plausible that it is phenomenologically viable. We have nothing new to add to this discussion either. It is however remarkable that a noncommutative spacetime provides a justification both for an arbitrary cosmological constant and for a quantized volume as well.

If nature is described by a canonical noncommutative spacetime, the only coordinate transformations compatible with the noncommutative algebra are four-volume preserving coordinate transformations. In that case, the cosmological constant is an arbitrary integration parameter and it is thus not related to a physical scale of the action.
This solves the first cosmological constant problem. More phenomenological arguments, well motivated by spacetime noncommutativity, lead to an estimate of today's value of the cosmological constant which agrees in order of magnitude with current fits to the astronomical data. This may be a  hint that quantum noncommutative gravity provides a solution to the second cosmological constant problem.
It is quite remarkable to find a relation between the cosmological constant i.e. dark energy and spacetime noncommutativity.  On a noncommutative spacetime, the size and evolution of the universe are intimately related to Planck scale physics.

\bigskip
\subsection*{Acknowledgments}
\noindent 
The author is grateful to Archil Kobakhidze for numerous and enlightening discussions on the cosmological constant within the framework of unimodular gravity.
He would like to thank Jean-Marie Fr\`ere, Steve Hsu, Philippe Spindel, Peter Tinyakov and Michel Tytgat  for helpful discussions. This work was supported in part by the IISN
(Belgian French community) and the Inter-University Attraction Pole V/27 (Belgian federal
science policy office).


\bigskip


\end{document}